\documentclass[%
 reprint,
 amsmath,amssymb,
 aps,
]{revtex4-2}

\usepackage{graphicx}% Include figure files
\usepackage{dcolumn}% Align table columns on decimal point
\usepackage{bm}% bold math
\usepackage{latexsym}
\usepackage{color}

\newcommand\Bigcircle{{\scalebox{1.5}{$\circ$}}}
\begin{document}

%\preprint{APS/123-QED}

\title{Probing the Gardner transition in an active quasi-thermal pressure-controlled granular system}% Force line breaks with \\

\author{Hongyi Xiao}%\email{hongyix@sas.upenn.edu}
\author{Andrea J. Liu}%\email{ajliu@physics.upenn.edu}
\author{Douglas J. Durian}%\email{djdurian@physics.upenn.edu}
\affiliation{%
 Department of Physics and Astronomy, University of Pennsylvania, Philadelphia, PA USA. 
}%

\date{\today}% It is always \today, today,

\begin{abstract}
To experimentally search for signals of the Gardner transition, an active quasi-thermal granular glass is constructed using a monolayer of air-fluidized star-shaped granular particles. The pressure of the system is controlled by adjusting the tension exerted on an enclosing boundary. Velocity distributions of the internal particles and the scaling of the pressure, density, effective-temperature, and relaxation time are examined, demonstrating that the system has important features of a thermal system. Using a pressure-based quenching protocol that brings the system into deeper glassy states, signals of the Gardner transition are detected via cage size and separation order parameters for both particle positions and orientations, offering experimental evidence of the Gardner transition for a system with in low spatial dimensions that is quasi-thermal.
%\begin{description}
%\end{description}
\end{abstract}

\maketitle

%\tableofcontents

%\section{\label{sec:level1}First-level heading:\protect\\ The line
%break was forced \lowercase{via} \textbackslash\textbackslash}

Glasses made of atoms, polymers, colloids, or grains, constitute an important class of materials, yet the understanding of their physics is far from complete~\cite{berthier2011theoretical,charbonneau2017glass}. In particular, some glasses are predicted to exhibit a Gardner transition, which is believed to occur as a stable glass is quenched into deeper glass states by lowering temperature or increasing pressure~\cite{gardner1985spin,berthier2019gardner}. At such a transition, the metastable basins in the energy landscape of a stable glass are further broken into a hierarchy of sub-basins, leading to a marginal glass phase with intermittency in its dynamics. Though the theoretical framework was established for spin glasses and structural glasses at high dimensions~\cite{gardner1985spin,gross1985mean,mezard1987spin,kurchan2012exact,kurchan2013exact,charbonneau2014fractal,parisi2020theory}, %as recognized in the 2021 Nobel Prize in Physics, 
the existence of the transition in two or three dimensions is still in question~\cite{urbani2015gardner,berthier2019gardner}.

A handful of studies have been carried out in search for the Gardner transition in lower dimensions using 
%idealized 
simulations 
%systems
~\cite{charbonneau2015numerical,berthier2016growing,scalliet2017absence,hicks2018gardner,li2021determining} and experiments~\cite{seguin2016experimental,hammond2020experimental,albert2021searching}. Various aspects of Gardner physics have been explored such as the cage statistics~\cite{charbonneau2015numerical,seguin2016experimental,hicks2018gardner}, aging and intermittent dynamics~\cite{hammond2020experimental,li2021determining}, and the growth of time and length scales~\cite{charbonneau2015numerical,berthier2016growing,li2021determining}. One strategy to detect the transition is to compress clones of a stable glass at an initial density into a higher density $\phi$. For $\phi<\phi_G$, where $\phi_G$ is the Gardner transition density, the clones can fully explore the metabasin of the glass. For $\phi>\phi_G$, the cloned configurations can be trapped in sub-basins, leading to a non-zero distance between these configurations that is larger than the average cage size~\cite{charbonneau2015numerical,seguin2016experimental,berthier2019gardner}. Experimentally, this strategy was first explored by cyclically compressing and relaxing a granular glass composed of vibrated circular disks~\cite{seguin2016experimental}. The results indicate a signal in the position-based cage order parameters for the Gardner transition.  However, a later study pointed out that possible cage-escaping events during quenching could complicate the interpretation of the results~\cite{hicks2018gardner}. 
%For non-idealized particles with friction or anisotropic shapes, particle orientations could be another important indicator for the Gardner transition, but this has not been explored. Furthermore, the Gardner transition is believed to occur over a narrow density range~\cite{seguin2016experimental}, but the corresponding pressure range should be much wider, which calls for a pressure-based detection protocol.

\begin{figure}[b]
\vspace{-5pt}
\includegraphics[width=7.5cm]{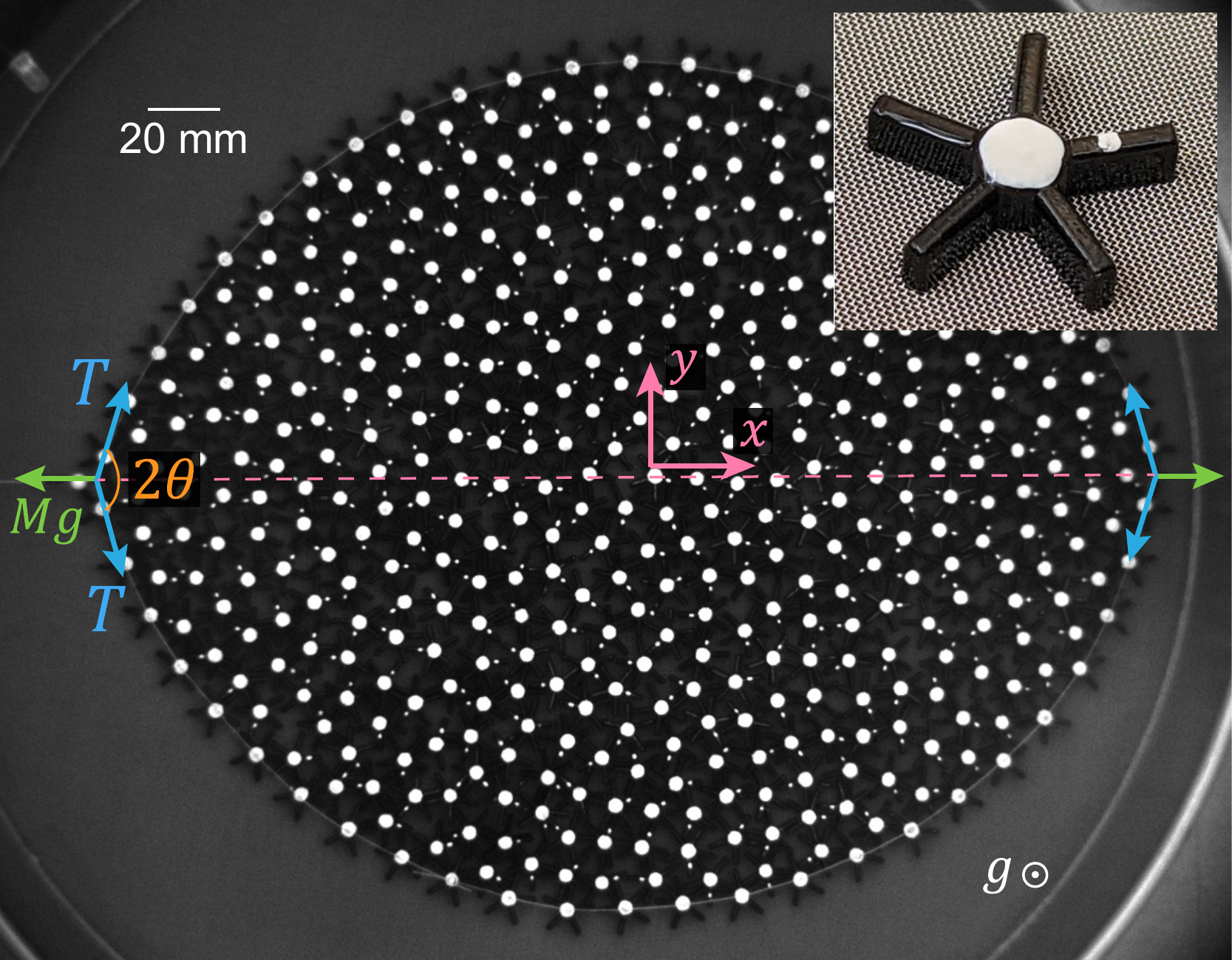}% Here is how to import EPS art
\caption{\label{fig:setup} Photograph of the pressure-controlled experiment with star-shaped particles (inset). Air-fluidized particles are enclosed by a boundary made of a chain of particles on a thin flexible thread. The left end is pulled with a controlled force $G=Mg$, while the right end is anchored.}
\end{figure}

In this letter, we present experimental evidence for the Gardner transition measured in a different kind of granular glass: a monolayer of air-fluidized star-shaped granular particles under controlled pressure. Three major differences exist in comparison to previous works. First, air-fluidization can inject energy with spatial and temporal uniformity into a granular layer by a sub-levitating upflow of air that sheds turbulent wakes to induce in-plane stochastic particle motion~\cite{ippolito1995granular,abate2005partition}. This energy injection scheme at the particle scale renders the system an active matter system and typically leads to particle behaviors that have a strong thermal analogy~\cite{daniels2012temperature,ojha2004statistical}. Unlike vibration-based approaches, air fluidization does not generate large convection currents of particles. Second, as the particle weight is largely countered by the air pressure, the basal friction between the particles and substrate becomes negligible. Third, The particles we use are designed to be star-shaped with five spokes (Fig.~\ref{fig:setup}). This allows us to explore the novel possibility that not just particle translations but also their orientations could be an important indicator of the Gardner transition. The five-fold symmetry prevents crystallization, and the concavity of the particles also enhances the caging of particles, which freezes the inherent structure at high pressure and prevents cage-breaking events from impairing signals of the Gardner transition. Fourth, the experiments are conducted under a pressure-controlled protocol, which distinguish our experiments from previous studies that mostly control density~\cite{charbonneau2015numerical,seguin2016experimental,hicks2018gardner}. This is an advantage because the Gardner transition is believed to occur over a narrow density range~\cite{seguin2016experimental}, but the corresponding pressure range should be much wider. 

We first show that the granular system exhibits quasi-thermal behavior, with Gaussian speed distributions and a well-defined equation of state. Then, using a pressure-based quenching protocol, we explore Gardner physics and demonstrate unambiguous signals in the caging statistics of both particle positions and orientation that indicate the appearance of energy sub-basins as the granular glass undergoes the Gardner transition.  

The particles are designed to be star-shaped with five evenly-spaced spokes surrounding a circular center with a diameter of 4.0~mm. The width and length of the spokes is 1.25~mm and 4.5~mm, respectively, defining a nominal diameter of $\sigma=13.0$\,mm. The height is 2.5~mm. The particles are 3D-printed with the Polyjet technique using a black rigid resin with a density of approximately 1.18~g/cm$^3$. The particles have mass $m=0.12$~g and moment of inertia $I=1.73$~g/mm$^2$ The experimental apparatus is based on a previously developed air-fluidization device~\cite{ojha2004statistical,abate2005partition,daniels2012temperature,farhadi2018dynamics}, where the particles are placed on a sieve with a mesh size of 150~$\mu$m. The air velocity is set to 3.6~m/s to minimize sliding friction between the particles and the sieve while not levitating the particles. 

As shown in Fig.~\ref{fig:setup}, the particles are enclosed by a flexible boundary consisted of 50 particles connected by a soft string with a center-to-center spacing of 14~mm. These boundary particles can also be fluidized, resulting in better uniformity of particle behaviors compared to rigid boundaries. The right end of the boundary is fixed while the left end is pulled by a constant force, $G=Mg$, where $M$ is the mass of a hanging wight and $g=9.8$~m/s$^2$. This results in a force balance of $G=2T\cos{\theta}$ at the ends with $T$ being the string's tension and $\theta$ being the half angle made by the string. For the rest of the boundary, a Laplace-like relation can be established for the pressure, $P$, exerted by the particles, which gives $P=T/R$, with $R$ being the radius of curvature. Thus, the pressure of the system is controlled and determined by the hanging mass via the relation $P=G/2R\cos{\theta}$.

Two series of experiments were performed. First, to study the thermal analogy and explore phase space, experiments with the system held at a constant pressure were conducted using various hanging weights $G$ and numbers $N$ of interior particles. In each case, the system is held at the desired $G$ for 10~min for equilibration, after which 1~min of data were taken. Second, to probe the Gardner physics, a pressure-based quenching protocol was developed, as described below. Videos of the experiments were recorded at 110~fps for the first set of experiments and at 12~fps for the quenching experiments, with a resolution of 50 pixels/$\sigma$. The center and one spoke of each particle were painted white to enable particle tracking~\cite{crocker1996methods}, with errors of around $0.004\sigma$ for position and 0.014~rad for orientation. Internal particles that are at least 3$\sigma$ away from the boundary are tracked for analysis, while boundary particles are used for calculating $\theta$, $R$, and the packing density $\phi$. A dynamical coordinate system is adopted with the $x$ direction aligned with the end-to-end vector.

\begin{figure}[b]
\vspace{-10pt}
\includegraphics[width=8.25cm]{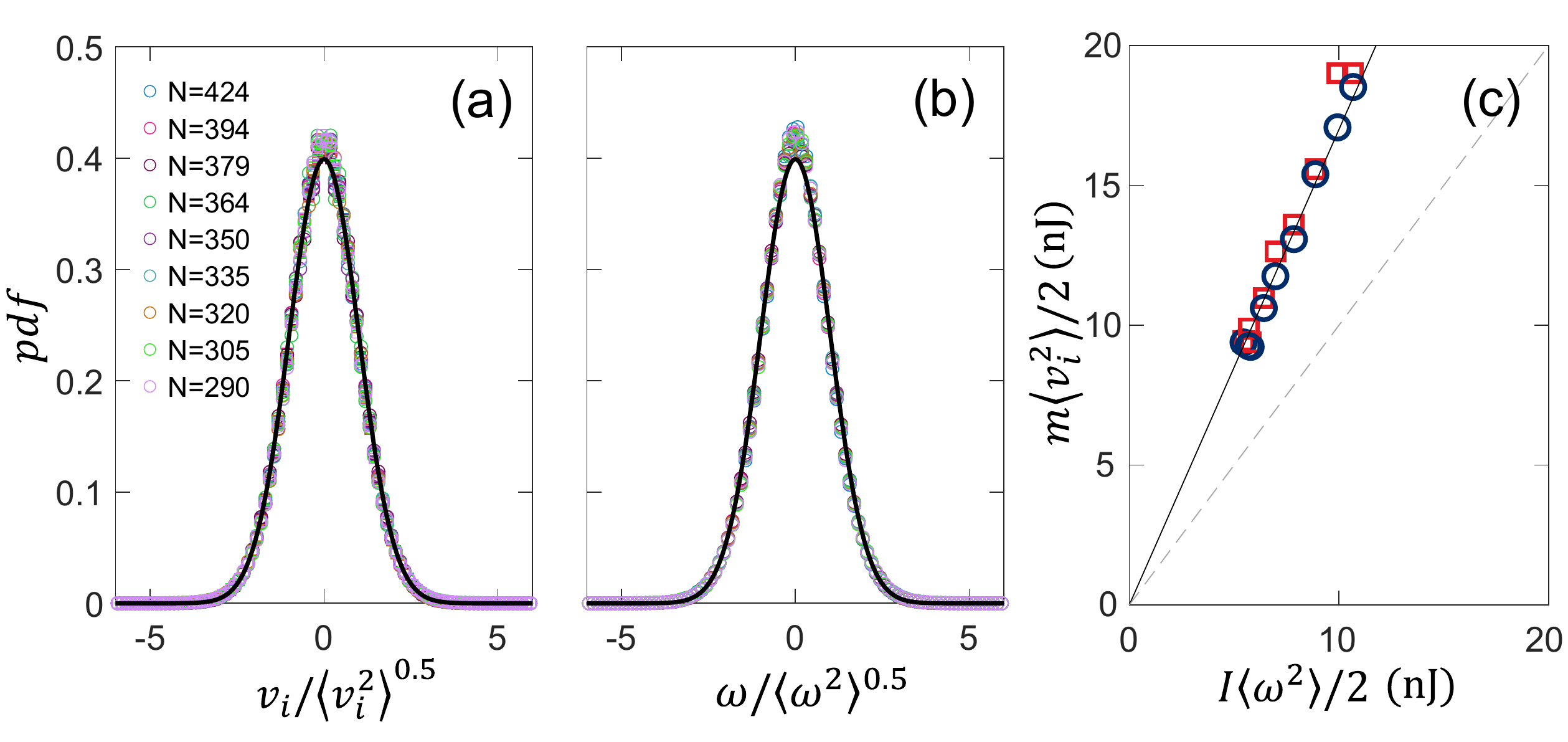}% Here is how to import EPS art
\vspace{-5pt}
\caption{\label{fig:vel} Distributions of (a) the normalized translational speeds, $v_i$, with $i=x,y$ and (b) the angular velocity, $\omega$, for internal particles from cases with different number of particles, $N$, under a constant hanging weight of $G=0.58\times 10^{-3}$~N. The black curves are the standard Gaussian distribution. (c) The translational energies $m\langle v_i^2 \rangle/2$ (red $\square$ for $i=x$, blue $\Bigcircle$ for $i=y$) vs the average rotational energy $I\langle \omega^2 \rangle /2$. The solid line is a linear fit with a slope of 1.7. }
\end{figure}

At low density and pressure, the behavior is strongly analogous to a liquid in thermal equilibrium. The distributions of translational and rotational velocities are shown in Fig.~\ref{fig:vel}a-b. Various cases are shown with $G=0.58\times 10^{-3}$~N kept constant and $N$ varied from $N=290$ to $N=424$. After normalizing the velocities by their root mean square value, all the data collapse to the standard Gaussian distribution, showing thermal behavior. The end-to-end distance fluctuates around a constant value, indicating mechanical equilibrium as well. Fig.~\ref{fig:vel}c examines the energy partition, showing that the two translational degrees of freedom have the same energy, $m\langle v_i^2 \rangle/2$, which is somewhat larger than the rotational energy, $I\langle \omega^2 \rangle /2$. The difference between the rotational and translational kinetic energies indicates that this active system is not perfectly thermal. %This suggests a possible decoupling between translation and rotation~\cite{edmond2012decoupling}.
In another set of such experiments conducted with $N=454$, and higher $G$ ranging from $0.58\times 10^{-3}$~N to $9.11\times 10^{-3}$~N, the system experiences a glass transition with the relaxation time greatly increasing; here, (quasi-)thermal equilibrium cannot be reached within a reasonable wait time. 

\begin{figure}[t]
\includegraphics[width=8cm]{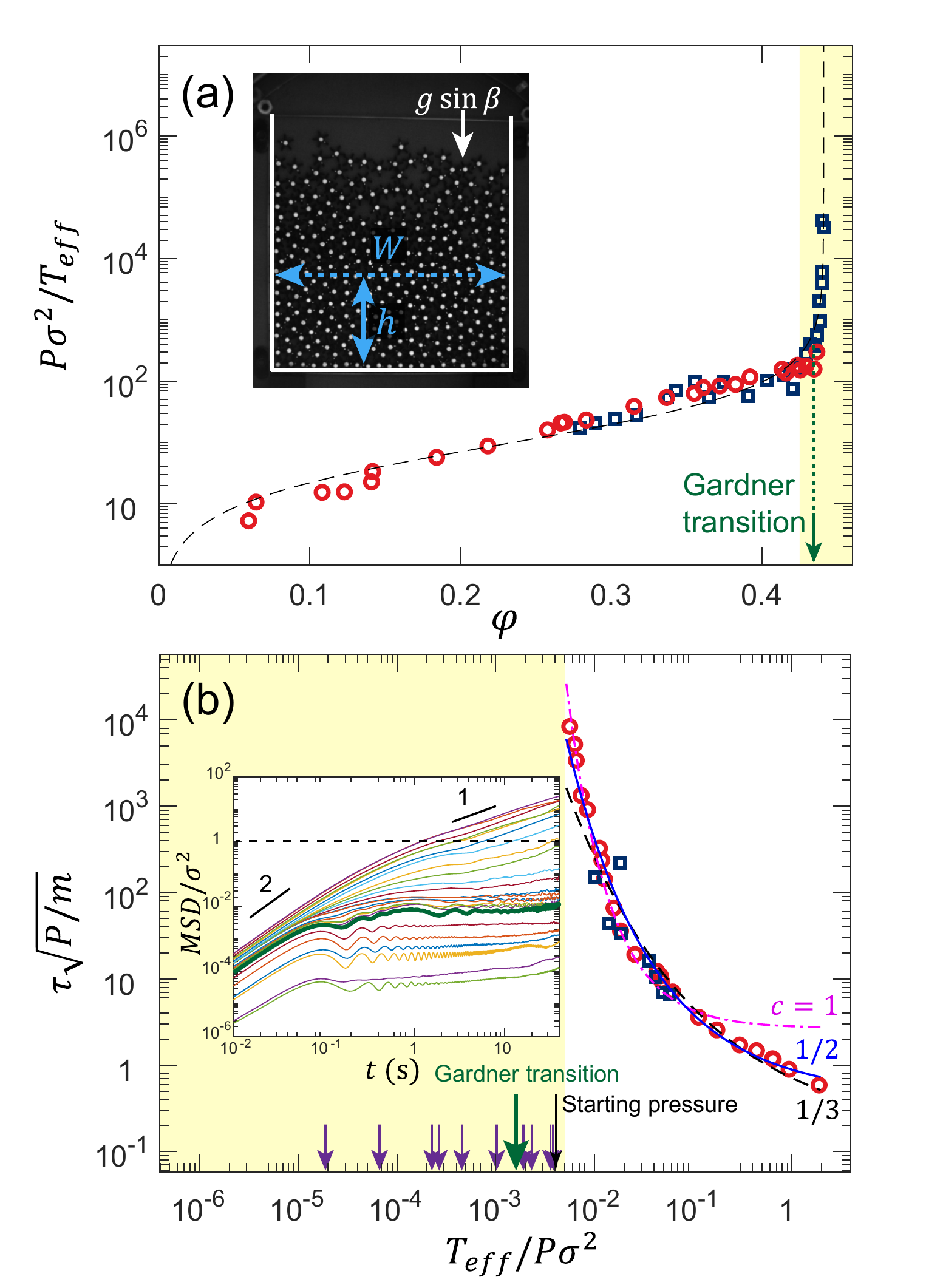}% Here is how to import EPS art
\vspace{-15pt}
\caption{\label{fig:eos} State parameters for the granular system: (a) Dimensionless pressure vs density for the pressure controlled (blue $\square$) and inclined (red $\Bigcircle$) experiments, with fitted equation of state (dashed curve). Inset shows the inclined experiment. (b) Dimensionless relaxation time vs dimensionless effective temperature for cases with measurable relaxation times. The curves represents fits to $y=a\exp(b/x^c)$ with $c$ as labeled. Inset shows the MSD for the pressure controlled experiments, with the thick green curve being the mean-squared displacement (MSD) at the Gardner transition. The shaded region marks the glass phase and the arrows represents the $T_{eff}/P\sigma^2$ of cases for testing the Gardner transition, with the corresponding density ranging from $\phi=0.429$ to $\phi=0.441$.}
\vspace{-10pt}
\end{figure}

To better explore the phase space, we performed auxiliary experiments by enclosing the particles in a rectangular box and tilting the apparatus by an angle of $\beta$~\cite{daniels2012temperature,farhadi2018dynamics} (inset of Fig.~\ref{fig:eos}a). In this way, the local hydrostatic pressure is determined by the depth, which gives $P=N_{a}(h)mg\sin{\beta}/W$ where $N_a(h)$ is the number of particles above a distance $h$ and $W$ is the width of the box. The corresponding local packing density, $\phi$, is extracted via binning. Several data points at different depths can be extracted from a single experiment, and various cases with varying $\beta$ and numbers of particles  were tested. 

Fig.~\ref{fig:eos}a examines the equation of state for the system using pressure, packing density, and an effective temperature defined from the translational energy of the particles, $T_{eff}=(1/2)m\langle v_x^2+v_y^2\rangle$. The dimensionless pressure, $P\sigma^2/T_{eff}$, is plotted verses $\phi$ for both the pressure-controlled and the inclined experiments, showing a good agreement and thus validating our pressure-control scheme. An equation of state based on a Free Volume Theory~\cite{kamien2007random,daniels2012temperature,farhadi2018dynamics} is fitted to the data, $P\sigma^2/T_{eff}\propto\phi/\left[1-(\phi/\phi_c)^{1/2} \right]$, where $\phi_c=0.441$ is approximately the point at which the pressure diverges and the jamming transition occurs. The fitted equation of state well describes the data for the system both in ``equilibrium" (corresponding to the experiments shown in Fig.~\ref{fig:vel}) and out of equilibrium~\cite{daniels2012temperature}, reinforcing the strength of the thermal analogy.

At $\phi$ close to $\phi_c$, the pressure increases drastically, indicating a glass phase. To identify the region of interest for exploring Gardner physics, the relaxation time, $\tau$, is examined, defined as the time for the mean squared displacement (MSD) of the particles to reach $\sigma^2$. Figure~\ref{fig:eos}b shows the dimensionless relaxation time, $\tau\sqrt{P/m}$, versus the dimensionless effective temperature, $T_{eff}/P\sigma^2$. The results can be best fitted to a stretched exponential~\cite{daniels2012temperature} with the form $\tau\sqrt{P/m}=a\exp\left[ b/ \left( T_{eff}/P\sigma^2 \right)^c \right]$, and the best fit is provided by $a=0.45$, $b=0.69$, $c=1/2$. Two reference curves are also shown with $a=0.14$, $b=1.62$, $c=1/3$, and $a=2.69$, $b=0.048$, $c=1$. As $T_{eff}/P\sigma^2$ decreases, the relaxation time increases rapidly and the MSD curve develops a plateau as a typical result of caging (Fig.~\ref{fig:eos}b inset). Per convention, we define the system to be in the glass phase when its inherent structure does not relax in 24~hours. Additional long-duration experiments were performed to determine the glass transition point, $P_g=0.012$~N/m and $\phi_g=0.425$, which corresponds to the experimental configuration with $N=454$, and $G=1.36\times10^{-3}$~N. Deep in the glass phase at even lower $T_{eff}/P\sigma^2$, the MSD slowly increases after the initial plateau, indicating aging as the system explores sub-basins, hinting at a Gardner transition~\cite{hammond2020experimental,berthier2016growing}. Fluctuations can be observed in the the initial plateau for these MSD curves; this is possibly another indicator of the transition.

To probe for the Gardner transition more quantitatively, we employ a quenching protocol in the glass phase (shaded region in Fig.~\ref{fig:eos}). which has a far wider pressure range than density range (about $0.03\phi_c$). We start with $G_0=1.55\times10^{-3}$\,N and quench the system to higher pressures in discrete steps. For each step, the weight is held at $G_0$ initially for 100~s, then instantaneously switched to a higher weight, $G_{high}$, and held for another 100~s for the system to approximately reach a mechanical equilibrium, setting the quenching pressure, $p_{high}$. Then $G$ is alternated between $G_0$ and $G_{high}$ for another ten cycles, with 20~s at $G_0$ and 10~s at $G_{high}$. We then move up to the next $G_{high}$, with a total of 11 values of $G_{high}$ tested, ranging from $1.65\times10^{-3}$~N to $8.72\times10^{-3}$~N, corresponding to the arrows labeled in Fig.~\ref{fig:eos}b. The short duration was selected to keep the total experiment time significantly shorter than the relaxation time, and to minimize aging in the marginal phase. 

\begin{figure}[t]
\includegraphics[width=8cm]{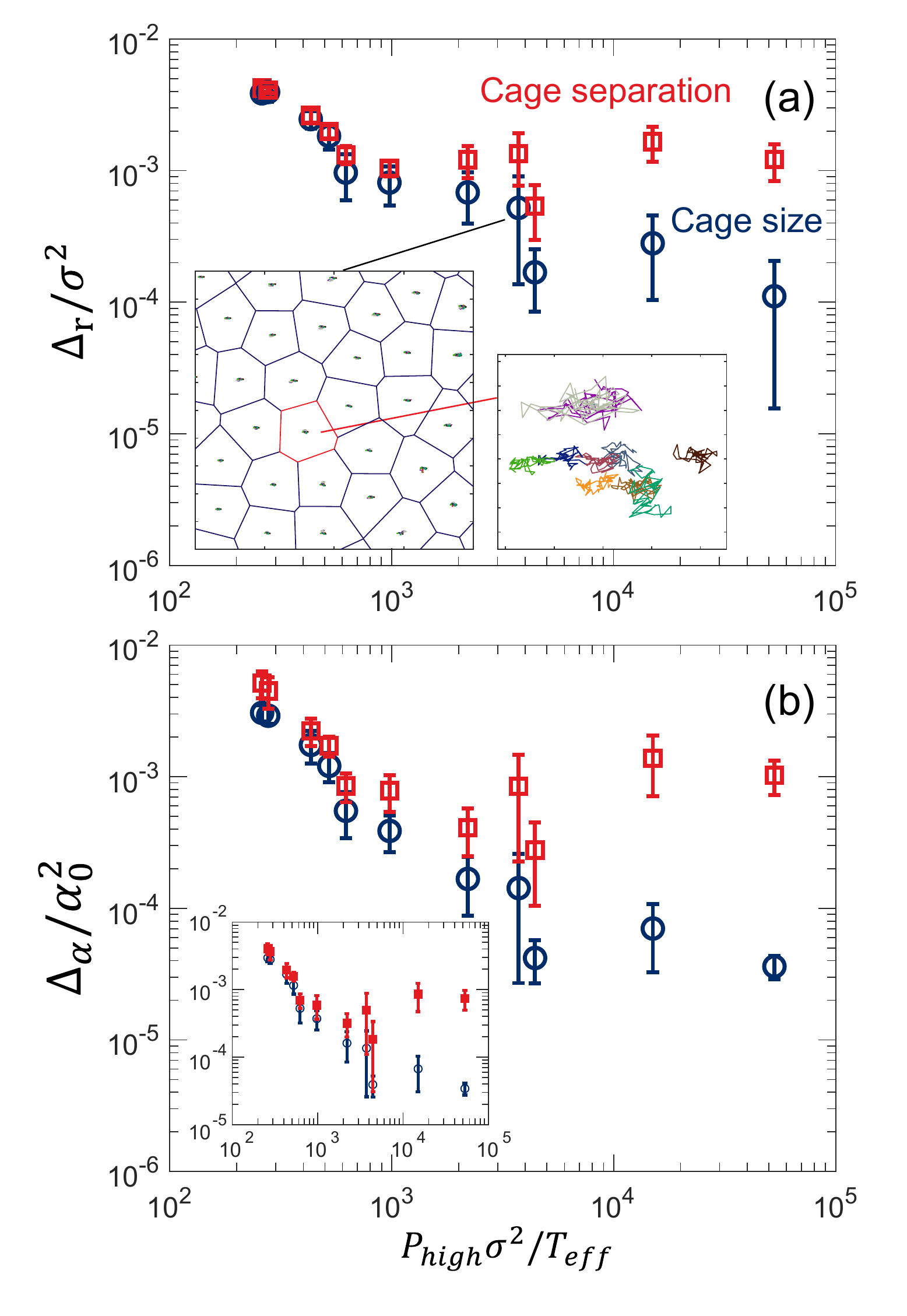}% Here is how to import EPS art
\vspace{-15pt}
\caption{\label{fig:gad} The transition to the Gardner phase is demonstrated by the onset of a difference between the cage separation $\Delta^{AB}$ (red $\square$) and the cage size $\Delta$ (blue $\Bigcircle$) order parameters with increasing $P_{high}$ normalized by the glass transition pressure $P_g$ for (a) positions and (b) orientations. The inset of (a) shows overlapping particle trajectories in the ten high pressure cycles along with the Voronoi cells for the initial cycle for $P_{high}\sigma^2/T_{eff}=3732$. The inset of (b) shows the cage order parameters with $\Delta^{AB}_{\alpha}$ calculated only using adjacent cycles. }
\vspace{-2pt}
\end{figure}

Following each pressure quench, we calculate the cage order parameters for particle positions, $\bold{r}_i$, as in previous studies~\cite{charbonneau2015numerical, seguin2016experimental, berthier2016growing}, and for particle orientations, $\alpha_i$. The first parameter is the characteristic cage size, which is based on the MSD in an individual quench, $k$, for both positions and orientations,
\vspace{-10pt}
\begin{equation}
    \begin{aligned}
    \Delta_r^k(t,t_w)=\frac{1}{\hat{N}}\sum_{i=1}^{\hat{N}} \lvert \bold{r}_i(t+t_w) - \bold{r}_i(t_w) \rvert^2, \\
    \Delta_{\alpha}^k(t,t_w)=\frac{1}{\hat{N}}\sum_{i=1}^{\hat{N}} \left( \alpha_i(t+t_w) - \alpha_i(t_w) \right)^2,
    \end{aligned}
\label{eq:self_r}
\end{equation}
where, $\hat{N}$ is the number of internal particles. Note that to minimize the effect of possible global deformation, we calculate $\bold{r}_i$ using the relative position of a particle to the centroid of its Voronoi neighbors~\cite{seguin2016experimental}. We set the wait time $t_w$ to be 5~s after the quench to $P_{high}$. Both $\Delta_r^k$ and $\Delta_\alpha^k$ quickly reach a plateau as particles are caged, and we define $\Delta_r=\langle \Delta_r^k \rangle$ and $\Delta_{AB}=\langle \Delta_\alpha^k \rangle$ as the average of the plateau values over all quenches.

For characterizing the cage separation between two quenches, $k$ and $k^\prime$, we calculate
\begin{equation}
    \begin{aligned}
    \Delta_r^{k,k^\prime}(t)=\frac{1}{\hat{N}}\sum_{i=1}^{\hat{N}} \lvert \bold{r}^{k}_i(t) - \bold{r}^{k^\prime}_i(t) \rvert^2, \\
    \Delta_{\alpha}^{k,k^\prime}(t)=\frac{1}{\hat{N}}\sum_{i=1}^{\hat{N}} \left( \alpha^{k}_i(t) - \alpha^{k^\prime}_i(t) \right)^2,
    \end{aligned}
\label{eq:self_r}
\end{equation}
using the positions and orientations of the same particle in the two quenches. We define $\Delta^{AB}_r=\langle \overline{ \Delta_r^{k,k^\prime}} \rangle$ and $\Delta^{AB}_\alpha=\langle \overline{ \Delta_\alpha^{k,k^\prime} } \rangle$ as averages over time (starting from the same $t_w$) and all combination of quenches. The order parameters are normalized by nominal cage sizes, which is $\sigma^2$ for position and $\alpha^2_0$ for orientation with $\alpha_0=2\pi/5$.

The results for the two cage order parameters calculated at each quench step are shown in Fig.~\ref{fig:gad}. At low $P_{high}$, the granular glass is in the stable glass phase where particles can fully explore their cages, resulting in no difference between the cage size $\Delta_r$ and the cage separation $\Delta_r^{AB}$. For the orientation order parameters at the first two $P_{high}$ points, $\Delta^{AB}_\alpha$ is slightly larger than $\Delta_\alpha$, which is likely the consequence of the global deformation that occurs as the system has not fully reached mechanical equilibrium. Calculating $\Delta^{AB}_\alpha$ by only comparing adjacent quench pairs reduces this difference, as shown in the inset to Fig.~\ref{fig:gad}.

As $P_{high}$ increases, the cage size $\Delta_r$ and $\Delta_\alpha$ decrease. While the cage separations $\Delta^{AB}_r$ and $\Delta^{AB}_\alpha$ follow the cage sizes at low $P_{high}$, they reach a plateau starting at $P_{high}\sigma^2/T_{eff} \approx 621$, defining the Gardner transition pressure, $P_G$ (labeled in Fig.~\ref{fig:eos}), with a corresponding density of $\phi_G=0.437$, which gives $\phi_G/\phi_c\approx0.99$, similar to the previously reported density ratio~\cite{seguin2016experimental}. For $P_{high}>P_G$, the difference between the size and the separation widens to larger than an order of magnitude for both position and orientation. This indicates that the granular glass is in an marginal phase, with each quench trapping the system in a different sub-basin~\cite{berthier2011theoretical, charbonneau2015numerical, seguin2016experimental}. In each $P_{high}$ step, the particles stay caged during the quench cycles, which can be seen from the example in Fig.~\ref{fig:gad} inset of the particle trajectories in all cycles, indicating that the difference between cage size and separation is not caused by cage-breaking as cautioned in a previous study~\cite{hicks2018gardner}. Particles are trapped in a sub-region of their cages during each quench, which can be clearly shown by an example of overlaid trajectories of an individual particle during the quench cycles with $P_{high}\sigma^2/T_{eff}=3732$.

In summary, the results in this study show significant, unambiguous experimental signals of the Gardner transition in both position and orientation, in an active quasi-thermal granular glass in two spatial dimensions. Our findings underscore the robustness of the Gardner transition, which was predicted in high dimensional thermal systems with only translational degrees of freedom. This study mainly focuses on using pressure quenches to probe sub-basins in the marginal glass, so aging~\cite{seguin2016experimental,hammond2020experimental} is avoided by minimizing the duration of the quenches. For a marginal glass held at a constant weight, the exploration of sub-basins in the energy landscape over longer duration could lead to experimental signals such as intermittent increasing of the end-to-end distance; this is the subject of future research. 

This work was supported by the National Science Foundation, grant MRSEC/DMR-1720530 (HYX) and by the Simons Foundation for the collaboration Cracking the Glass Problem Award 454945 (A.J.L) and Investigator Award 327939 (A.J.L.).

%\nocite{*}
\def\bibsection{\section*{\refname}}
\bibliography{gardner.bib}% Produces the bibliography via BibTeX.

\end{document}